\documentstyle[preprint,aps]{revtex}
\def\be{\begin{equation}}
\def\ee{\end{equation}}
\def\bea{\begin{eqnarray}}
\def\eea{\end{eqnarray}}

\begin{document}

\title{The Gross-Neveu Model with Chemical Potential; An Effective Theory 
for Solitonic-Metallic Phase Transition in Polyacetylene?}

\author{Hisakazu Minakata\footnote[2]{Talk presented at APCTP-ICTP Joint 
Conference '97 (AIJIC97) on Recent Developements in Nonperturbative 
Quantum Field Theory, May 26 - 30, 1997, APCTP, Seoul, Korea}}

\address{Department of Physics, Tokyo Metropolitan University \\
1-1 Minami-Osawa, Hachioji, Tokyo 192-03 Japan}

\author{Alan Chodos}

\address{Center for Theoretical Physics, Yale University \\
217 Prospect Street, New Haven, CT 06511-8167 USA}

\date{September, 1997}

\preprint{
\parbox{5cm}{
TMUP-HEL-9708\\
YCTP-P18-97\\
hep-th/9709179\\
}}

\maketitle

\begin{abstract}
The Gross-Neveu model with chemical potential is investigated as a 
low-energy effective theory of polyacetylene. In particular, we focus 
on the abrupt change in the features of electric conductivity such as 
sharp rise in the Pauli paramagnetism at dopant concentration of about 
6\%. We will try to explain it by the finite density phase transition 
in the Gross-Neveu model. The thermodynamic Bethe ansatz is combined 
with the large-N expansion to construct thermodynamics of the 
Gross-Neveu model. A first-order phase transition is found in leading 
order in the 1/N expansion and it appears to be stable against the 1/N 
correction. The next to leading order correction to the critical dopant 
concentration is explicitly calculated.   
\end{abstract}
\newpage

\section{What is interesting in electric conductivity in polyacetylene?}

In this talk we want to describe our recent works on thermodynamics of 
the Gross-Neveu model with chemical potential.\cite{CM1,CM2}
The physical motivation for these works comes from condensed matter 
physics; electric conductivity properties of doped polyacetylene.\cite{HKSS}
In particular, we are trying to describe the abrupt change in electric 
conductivity properties at dopant concentration of about 6\%. 

Polyacetylene is a typical 1-dimensional polymer and we are particularly 
interested in trans-type polyacetylene. It is a fascinating material, 
showing variety of features of electric conductivity depending upon the 
dopant concentration.\cite{HKSS} At zero doping it is an insulator. 
Because of the Peierls instability an electron gas in 1 dimension is 
unstable against generation of charge density wave \cite{Kittel} 
which entails, in this system, an alternating distortion of the lattice. 
It is called dimerization in condensed-matter literatures and the ground 
state is doubly degenerate. As a consequence an energy gap opens above 
the Fermi sea which implies an insulator.  

At dopant concentration up to $\sim$6\% the electric conductivity grows 
as dopant concentration increases but the Pauli paramagnetism stays low, 
indicating that the charge carrier is spinless. 
(A note for non-condensed matter physicists; doping is a procedure of 
importing the impurity atoms into the material to supply extra electrons 
or holes into the system. The dopant concentration is defined as the 
number of doped electrons per carbon atom.)
It is believed that electric charge is transported by the spinless charged
solitons in polyacetylene in this intermediate doping region.\cite{HKSS} 
One can say that the system displays the particular type of charge-spin 
separation.

It appears that there is a consensus that the relevant dynamical degrees 
of freedom are solitons in this regime. Their existence is in fact 
predicted by Su, Schrieffer and Heeger in their original paper.\cite{SSH} 
They introduced the SSH Hamiltonian to describe polyacetylene and discussed 
its physical implications. The spectrum of solitons in this 
model contains two spinless charged solitons and two neutral solitons 
with spin 1/2. Prior to their pioneering work it was observed by Jackiw 
and Rebbi \cite{JR} that in theories with fermions solitons can carry 
fractional fermion quantum numbers. The peculiar features of spectrum 
of solitons in the SSH model can be nicely explained by the phenomenon 
of fermion fractionization.

Then, at dopant concentration of about 6\% an abrupt rise occurs in 
the Pauli paramagnetism while the conductivity keeps monotonically 
increasing. In some samples the conductivity reaches as high as that 
of copper.\cite{Tsuka}
It is likely that the behavior indicates a transition to a metallic phase. 
This is the behavior that we want to understand. 

\section{The Gross-Neveu model as a low-energy effective theory of 
polyacetylene}

Soon after the proposal of the SSH Hamiltonian, it was shown by 
Takayama, Lin-Liu and Maki \cite{TLM} that the low-energy effective 
continuum theory of the SSH model is given by a 1+1 dimensional 
relativistic four-Fermi theory with two-flavor Dirac fermions. 
The Lagrangian of the model can be written by using an auxiliary field 
$\sigma$ (which represents the phonon degrees of freedom) and suppressing 
the flavor index as
\begin{equation}
{\cal L} = \overline{\psi} i \gamma^{\mu} \partial_{\mu} \psi - 
\frac{1}{2}{\sigma}^2 - g\sigma\overline{\psi}\psi
\end{equation}
This is nothing but the field theory model which was first investigated 
by Gross and Neveu \cite{GN} as an asymptotically-free renormalizable   
quantum field theory with many common features with quantum 
chromodynamics. 

We would like to utilize the Gross-Neveu model with chemical potential 
as the effective theory for a transition from solitonic to metallic phase. 
If it does a good job the model should display a phase transition at a 
certain value of the chemical potential. Is it likely? 
We believe that the answer is yes. It is believed that at zero chemical 
potential the ground state of the model violates the discrete chiral 
symmetry and it is doubly degenerate. 
It is, of course, in accord with the dimerized ground state and 
is nothing but the manifestation of the Peierls instability. Therefore, 
the theory admits solitons (kinks) and they are ``fermion-fractionized'' 
solitons in the sense of Jackiw and Rebbi.\cite{JR} 
The evidence for the chiral symmetry breaking and the doubly degenerate 
ground state comes from the large-N approximation.\cite{GN} It is also 
strongly supported by the existence of consistent 
factorizable fermion-fermion,\cite{Zamo1} fermion-kink and kink-kink 
S-matrix.\cite{SW,KT} 

At large enough $\mu$, on the other hand, we believe that the symmetry is 
restored. Because the Gross-Neveu model is an asymptotically free theory 
the fermions become non-interacting at high enough densities and are not 
able to support their condensate that they formed to break chiral symmetry 
at low densities. Therefore, it is very likely that the theory displays 
the phase transition at a certain fermion density or at the corresponding 
value of chemical potential. We note that this is in disagreement with 
what the soliton lattice theory predicts.\cite{Horo}   

\section{The Gross-Neveu model with chemical potential; the large-N limit}

To verify (or disprove) this conjecture and to know the order of the 
phase transition (if any) we examined the Gross-Neveu model with chemical 
potential by using the large-N approximation.\cite {CM1} 
In the leading order of 1/N expansion the effective potential of the 
model can be obtained analytically by taking into account the effect 
of chemical potential. It reads: 

\begin{eqnarray}
\lefteqn{V_{eff}(\sigma)}\nonumber\\
&& = \frac{\sigma^2}{2N} + \frac{\sigma^2}{4N} \left[\theta  
(\sigma^2 - \gamma^2) \{ln (\sigma^2/\sigma_{0}^{2}) - 3\} +  
\theta(\gamma^2 - \sigma^2) \{2 ln 
\frac{\gamma + \sqrt{\gamma^2 -  \sigma^2}}{\sigma_0} -3 \}\right] \nonumber\\
&& - \frac{\gamma}{2N} \sqrt{\gamma^2 - \sigma^2} \theta(\gamma^2 -  
\sigma^2) 
\label{potential}
\end{eqnarray}
where $\gamma=\mu\sqrt{N/\pi}$. 
By analyzing the expression one can easily figure out that at 
$\mu=m/\sqrt{2}$, where $m$ is the mass of the fundamental fermion, 
the theory has a triple degenerate ground states. 
(See Ref. 1 for details.) 
This is a clear signature for the first-order phase transition. One can 
evaluate the corresponding value of the fermion density, which can be 
interpreted as dopant concentration, as 
\[
y_c = \frac{N}{\sqrt{2}\pi\xi} 
\]
where $\xi$ is the correlation length (soliton size) measured in units of 
lattice constant. It is estimated as $\xi \simeq 7$ by SSH.\cite{SSH} 
It turns out that $y_c = 0.064$, very close to 6\% at $N=2$. 
Here $N$ is the number of flavor of Dirac fermion in the effective 
field theory. The two flavor originates the spin degrees of freedom of the 
electrons in the SSH model. 

\section{Going beyond the large-N limit}

It is a good news that the large-N works but there arise a number of 
questions. 
\begin{enumerate}
\renewcommand{\labelenumi}{(\arabic{enumi})}
\item 
Are there any experimental evidences for the first-order phase transition?
\item
Is the large-N approximation reliable even at N=2?
\item
Can the result be viewed as a robust evidence for the presence of finite 
density phase transition in the Gross-Neveu model? 
\end{enumerate}
Again a good news about (1) is that apparently there exists an experiment 
\cite{CCMH} that signals a hysteresis in the Pauli paramagnetism as a 
function of chemical potential; 
a clear signature for first-order phase transition. 
The result, however, seems to be neither confirmed nor refuted by 
the other independent experiments. 

The second and the third questions are more pressing to us as theorists. 
Certainly we should be able to do better. As a first step toward answering 
these questions we have calculated the next to leading order correction 
to see if the first-order phase transition survives 
(we mean, if there are any instabilities occurring) and to compute 
correction to the critical dopant concentration.\cite{CM2} 

To carry this out, however, we had to invent a novel way of computation 
because the direct calculation of the effective potential at two-loop 
with chemical potential looks to be formidable. The method we employed 
is a hybrid method combining the thermodynamic Bethe ansatz (TBA) 
and the 1/N computation at zero chemical potential. 

\section{The thermodynamic Bethe ansatz}

The thermodynamic Bethe ansatz \cite{YY,Zamo2} is a powerful method 
for constructing thermodynamics of a 1 dimensional system of particles 
whose $S$-matrix elements $S_{ab}$ are exactly calculable. The technique 
is particularly suited to our problem because all the $S$-matrix elements 
of the Gross-Neveu model are known.\cite{Zamo1,SW,KT}
By solving the integral equation of TBA one can, in principle, construct 
an exact thermodynamics of the Gross-Neveu model. 

Let us restrict ourselves to the elastic (or diagonal) scattering 
theories in which neither inelastic nor ``charge-exchange'' reactions 
occur.\cite{KM} Following Yang and Yang \cite{YY} one can write 
down the integral equation for dressed energy $\epsilon_a (\theta)$ of 
quanta as a function of the rapidity $\theta$, where the subscript $a$ 
specifies the particle species. It reads, 

\begin{equation}
T \epsilon_a(\theta) = -\mu_a + m_a\cosh \theta 
- i T\sum_b \int d{\theta}'
K_{ab}(\theta- {\theta}') 
ln \left[ 1+ \mbox{e}^{-\epsilon_b({\theta}')}\right]
\end{equation}
where the kernel $K_{ab}(\theta)$ is defined by
\begin{equation}
K_{ab}(\theta) = ({1 \over 2\pi i}) {d ln S_{ab}(\theta) \over d\theta} 
\end{equation}
Having solved the equation for $\epsilon_a (\theta)$ the free energy 
density of the system can be computed as 

\begin{equation}
f(\mu) = -\sum_a\int d\theta m_a\cosh\theta ln
\left[1+\mbox{e}^{-\epsilon_{a}}\right] + \sum_a\mu_a D_a
\label{freeenergy}
\end{equation}
where $D_a$ denotes the number density of particle $a$. 
We note that (negative of) the first term in (\ref{freeenergy}) implies 
the pressure $P$ because of the thermodynamic relation $f = \mu D - P$.  
Hereafter, 
we consider the case of single species of particles. We argue that the  
approximation can be justified in the large-N expansion of the 
Gross-Neveu model. 

Since we are interested in the possibility of phase transitions at 
finite chemical potential we want to take zero temperature limit. 
It is because there is no phase transition at finite temperature in 
1+1 dimensions due to the Mermin-Wagner theorem.\cite{MW}
\footnote{
One can argue in length that the the Gross-Neveu model at zero-temperature 
serves as a better effective theory of pure sample of polyacetylene at 
the room temperatures. The Mermin-Wagner theorem holds because of a 
strict one-dimensionality of space and it is known that its prohibition 
becomes invalid upon turning on very tiny three-dimensionality such as 
infinitesimal interchain couplings. H. M. thanks Yutaka Okabe for 
discussion on this point.}
One can show that by taking zero-temperature limit $T\rightarrow 0$ 
with $T\epsilon(\theta) \equiv - \tilde{\epsilon}(\theta)$ kept finite, 
the TBA equation for ``energy density'' takes the form, \cite{FNW} 

\begin{equation}
\tilde{\epsilon}(\theta) = \mu - m cosh\theta + \int_{-B}^{B}
d\theta^{\prime} K(\theta - \theta^{\prime})
\tilde{\epsilon}(\theta^{\prime}).
\end{equation}
where $B$ is determined by $\tilde{\epsilon}(B)=0$ and the integration 
region is restricted to $\tilde{\epsilon}(B) > 0$. 
Using the solution of this integral equation the pressure can be 
expressed in zero-temperature limit as 
\begin{equation}
P = (\frac{m}{2\pi}) 
\int_{-B}^{B}d\theta cosh\theta \tilde{\epsilon}(\theta) 
\end{equation}
The negative of the pressure is the relevant quantity to discuss the 
finite-density phase transition and to calculate the critical dopant 
concentration. It is related with the grand canonical partition function 
$\Xi$ as $-PL = - T ln\Xi$.\cite{Kubo}
(Recall the similar relationship $F = -T ln Z$ between the Helmholtz 
free energy $F$ and the canonical partition function Z.)
The fermion number density n is then given by $- \frac{dP}{d\mu}$. 

There are, however, obstacles both technical and conceptual. 
The technical problem is that the integral equation is difficult to 
solve analytically. But it can be done numerically. On the other hand 
the conceptual problem is more difficult to handle. We do not know how 
the phase transition can be signalled in the framework of the TBA. 

\section{The TBA in leading order in 1/N expansion}

To gain insight to this problem we have examined the question by using 
the 1/N expansion.\cite{CM2} At large N one can argue that the only 
relevant degrees of freedom are fundamental fermions because everybody 
else is more massive, with the masses being proportional to N. 
Therefore, the single species approximation should be valid. 
We have found an interesting structure. 
In leading order of 1/N expansion we have the analytic 
expression (\ref{potential}) of the effective potential of the 
Gross-Neveu model with chemical potential. 
From this we can obtain the absolute value of the ground state energy 
density as a function of chemical potential:
\begin{equation}
V_{eff} = - {m^2 \over 4\pi} + \theta(\mu^2 - m^2) {m^2 \over 2\pi}
[ln {\mu + \sqrt{\mu^2 - m^2} \over m} - {\mu \over m} \sqrt{{\mu^2
\over m^2} - 1}].
\label{Vmin}
\end{equation}
Here we have chosen the renormalization point $\sigma_0$ so that 
$\lambda \equiv g^2N = \pi$ and then $\sigma_0$ gives minimum of the 
effective potential.  
 
On the other hand, we can compute the free energy of the model by 
solving the TBA equation. In leading order of 1/N it is trivial to 
solve the equation because the kernel $K(\theta)$ vanishes 
(namely, $S=1$). We obtain for (negative of) the pressure, $-P$,
the exactly the same expression as the second term in (\ref{Vmin}).
The pressure given by the TBA is normalized such that it vanishes 
at $\mu=m$ and there is no way (to our knowledge) of computing it below
$\mu=m$. 

It is conceivable that the pressure $-P(\mu)$ has additional 
$\mu$-independent contribution $-P(0)$. It can be interpreted as
the vacuum energy density of the Gross-Neveu model. On dimensional 
ground it can be written as $-P(0) = -bm^2$, where $b$ is a constant.  
While there is no way of computing $b$ within the framework of the TBA 
it is calculable, at least in principle, in the 1/N expansion. 
In the leading order it was computed by Gross and Neveu \cite{GN} and 
$b=1/4\pi$. Together with the TBA result of $-P(\mu)$, it 
reproduces the exact expression of the free energy, i.e., the 
effective potential (\ref{potential}) 
obtained by the explicit 1/N calculation. 

This establishes our strategy of how to compute free energy of the 
Gross-Neveu model order by order in 1/N expansion. Namely, we obtain 
$-P(\mu)$ by solving the TBA equation and supplement it by 
the vacuum energy density calculated at zero chemical potential. 

Then, the question is how the TBA free energy can signal the phase 
transition? We propose to take massless limit of the pressure 
$-P(\mu)-P(0)$ 
calculated by the above method to obtain the pressure in the 
massless phase.\cite{CM2} We have 
\begin{equation}
-P_{massless}(\mu) = - \frac{\mu^2}{2\pi}, 
\end{equation}
the free fermion result. Then, we ask if the two pressures 
cross at certain value of the chemical potential. They do at 
$\mu=m/\sqrt{2}$, reproducing the phase transition point predicted 
by the explicit leading large-N computation.

\section{The TBA in next to leading order}

In our second paper we have gone one step further by generalizing the 
procedure to the next to leading order in 1/N expansion.\cite{CM2}
The TBA equation is no longer trivial to solve but fortunately it can 
still be solved in a closed form. 
We skip all the technicalities and just quote the result.\cite{CM2,FNW} 
The 1/N correction to the free energy of the Gross-Neveu 
model can be given in a parametric form:
\begin{equation}
-P(\mu)  = - \frac{m^2}{N\pi}[B^2 + sinh^2 B - BShi(2B)] 
\end{equation}
\begin{equation}
\mu = m cosh B + 
\frac{1}{N}\left[
m cosh B[Shi(2B) - 2B] - m sinh B[Chi(2B) - ln(sinh 2B) - \gamma]
\right]
\end{equation}
where $Shi(x)$ and $Chi(x)$ are the hyperbolic integral functions
defined by
\begin{eqnarray}
&& Shi(x) \equiv \int_{0}^{x} dt {sinh t \over t}, \nonumber\\
&& Chi(x) \equiv \gamma + lnx + \int_{0}^{x} dt {cosh t - 1 \over t}.
\end{eqnarray}

The uncalculable constant, the vacuum energy density  
at zero chemical potential, was indeed calculated long time ago by 
Sch\"onfeld.\cite{Sch} With a bit of numerical computation the 
correction to the constant $b$ can be written as
\begin{equation}
b = {1 \over 4\pi}[1 - {2.12 \over 3N}].
\end{equation}
Then, one can 
go through the exactly the same procedure as the leading order. 
We still obtain the free fermion result by taking the massless limit 
but with chemical potential with 1/N correction. It is what we expect 
because of the asymptotic freedom of the Gross-Neveu model. 
(See Ref. 2 for details.)
The resulting correction to the critical point can be written as 
\[
\mu_c = \frac{m}{\sqrt{2}}[1-\frac{0.47}{N}].
\]
Therefore, it gives rise to about 20\% correction to $\mu_c$ at N=2. 
If it is expressed by critical dopant concentration $y_c = 0.05$, 
which is somewhat smaller than but still roughly agrees with the  
experimental value.  

\section{Conclusion and outlook}

In this talk we have discussed how far one can proceed toward the 
thermodynamics of the Gross-Neveu model at zero temperature and 
finite background fermion density. At least in the 1/N expansion 
we have found that there is persistent first-order phase transition 
at a certain value of the chemical potential. Whether it survives 
in an exact treatment at N=2 theory remains to be seen. However, 
we have argued that it is the case by relying on the following 
robust two features of the model: 
one is the chiral symmetry breaking at zero chemical potential 
as manifested in the kink spectrum which is known exactly by the 
method of factorizable S-matrix, and the other, the asymptotic 
freedom of the theory.

Is it possible to go beyond the large-N expansion? It is a very 
nontrivial question. In principle the answer is yes. Since all 
the S-matrix elements are known it should be possible to formulate 
the complete thermodynamics of the Gross-Neveu model. A more difficult 
but challenging question is whether the phase transition can be 
signaled solely within the framework of the TBA. We are now trying 
to answer (at least some of) these questions.

\vspace*{-2pt}
\section*{Acknowledgments}
This work was performed as an activity under support by Agreement 
between Tokyo Metropolitan University and Yale University on Exchange 
of Scholars and Collaborations, engaged in May 1996. 
H.M. is partially supported by Grant-in-Aid for Scientific Research 
\#09640370 of the Ministry of Education, Science and Culture, 
and by Grant-in-Aid for Scientific Research \#09045036 under 
International Scientific Research Program, Inter-University 
Cooperative Research. A.C. is supported in part by the DOE under grant 
number DEFG0292ER40704. 


\end{document}